\documentclass{article}

\def\xxinput#1{\input#1}

\xxinput{vsolj02.sty}

\usepackage{longtable}
\usepackage[dvipdfmx]{graphicx}
\usepackage[comma,colon]{natbib}

\usepackage[OT2,T1]{fontenc}

\def\cite{\citealt}
\setcitestyle{aysep={}}

\newcounter{author}
\def\altaffilmark#1{$^{#1}$}
\def\altaffiltext#1{$^{#1}$\,}

\def\authorcount#1#2{{\refstepcounter{author}\label{#1}
                     \altaffiltext{\ref{#1}}{#2}}}

\begin{document}

\begin{center}

\title{Long-lasting standstill and fading episode in the IW And star V507 Cyg}

\author{
        Taichi~Kato,\altaffilmark{\ref{affil:Kyoto}}
        Masayuki~Moriyama\altaffilmark{\ref{affil:Myy}}
}
\email{tkato@kusastro.kyoto-u.ac.jp}

\authorcount{affil:Kyoto}{
     Department of Astronomy, Kyoto University, Sakyo-ku,
     Kyoto 606-8502, Japan}

\authorcount{affil:Myy}{
     Variable Star Observers League in Japan (VSOLJ),
     290-383, Ogata-cho, Sasebo, Nagasaki 858-0926, Japan}

\end{center}

\begin{abstract}
\xxinput{abst.inc}
\end{abstract}

   V507 Cyg is one of the prototypical IW And-type stars
\citep[for references of IW And stars, see][]{sim11zcamcamp1,szk13iwandv513cas,ham14zcam,kat19iwandtype}.
We report on the detection of a long standstill in
2020 May--2022 March.  We used observations by
the All-Sky Automated Survey for Supernovae (ASAS-SN)
Sky Patrol data \citep{ASASSN,koc17ASASSNLC},
the Zwicky Transient Facility (ZTF: \cite{ZTF})
data\footnote{
   The ZTF data can be obtained from IRSA
$<$https://irsa.ipac.caltech.edu/Missions/ztf.html$>$
using the interface
$<$https://irsa.ipac.caltech.edu/docs/program\_interface/ztf\_api.html$>$
or using a wrapper of the above IRSA API
$<$https://github.com/MickaelRigault/ztfquery$>$.
} and unfiltered snapshot CCD observations by
one of the authors (MM).

\begin{figure*}
\begin{center}
\includegraphics[width=16cm]{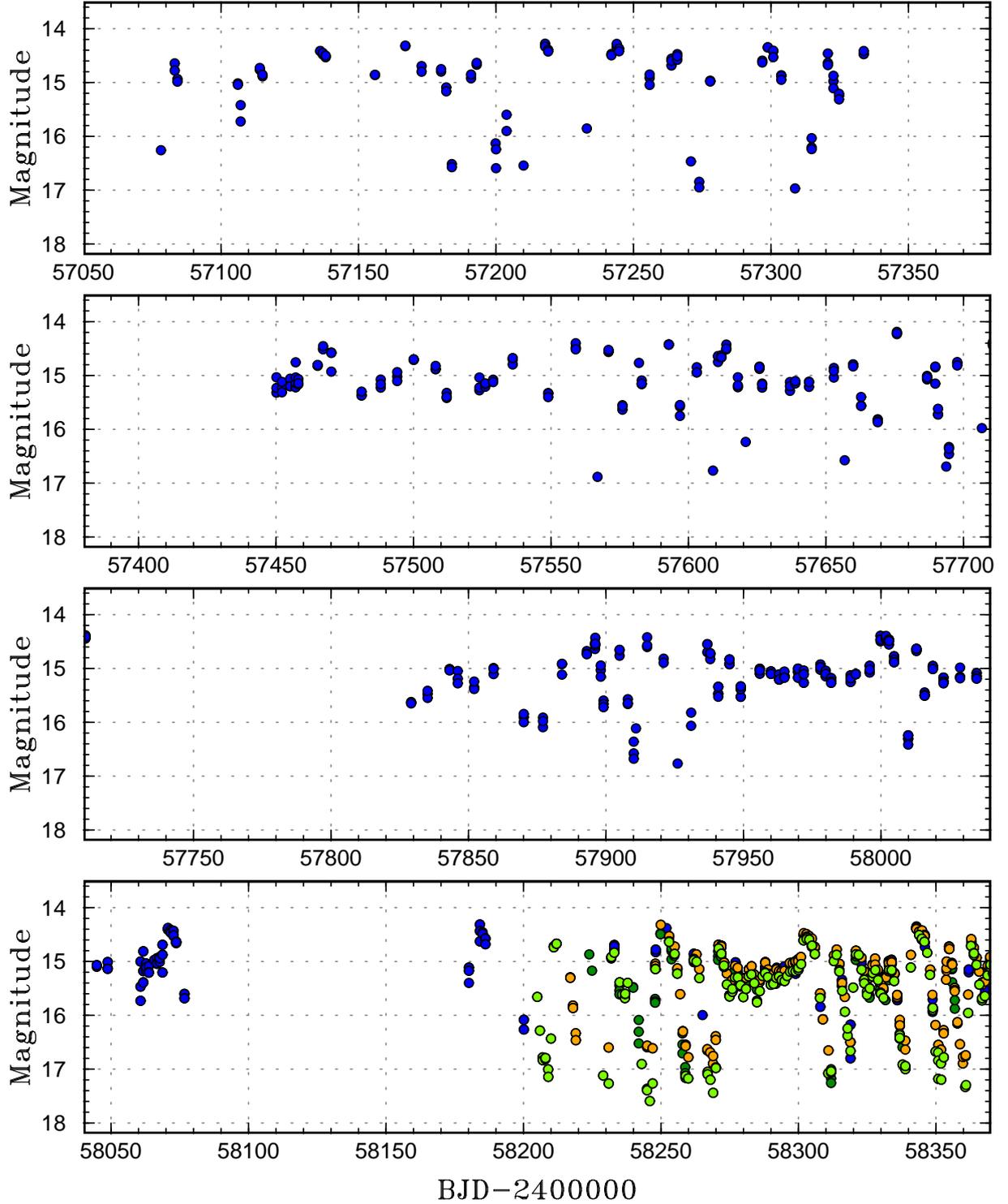}
\caption{
  Long-term light curve of V507 Cyg using
  ASAS-SN and ZTF observations.
  See figure \ref{fig:v507cyglc2} for the symbols.
  V507 Cyg primarily showed dwarf nova-type
  variations sometimes interrupted by short standstills
  bearing characteristics of IW And stars.
  This figure corresponds to figure 2 in
  \citet{kat19iwandtype} with subsequent observations.
}
\label{fig:v507cyglc}
\end{center}
\end{figure*}

\begin{figure*}
\begin{center}
\includegraphics[width=16cm]{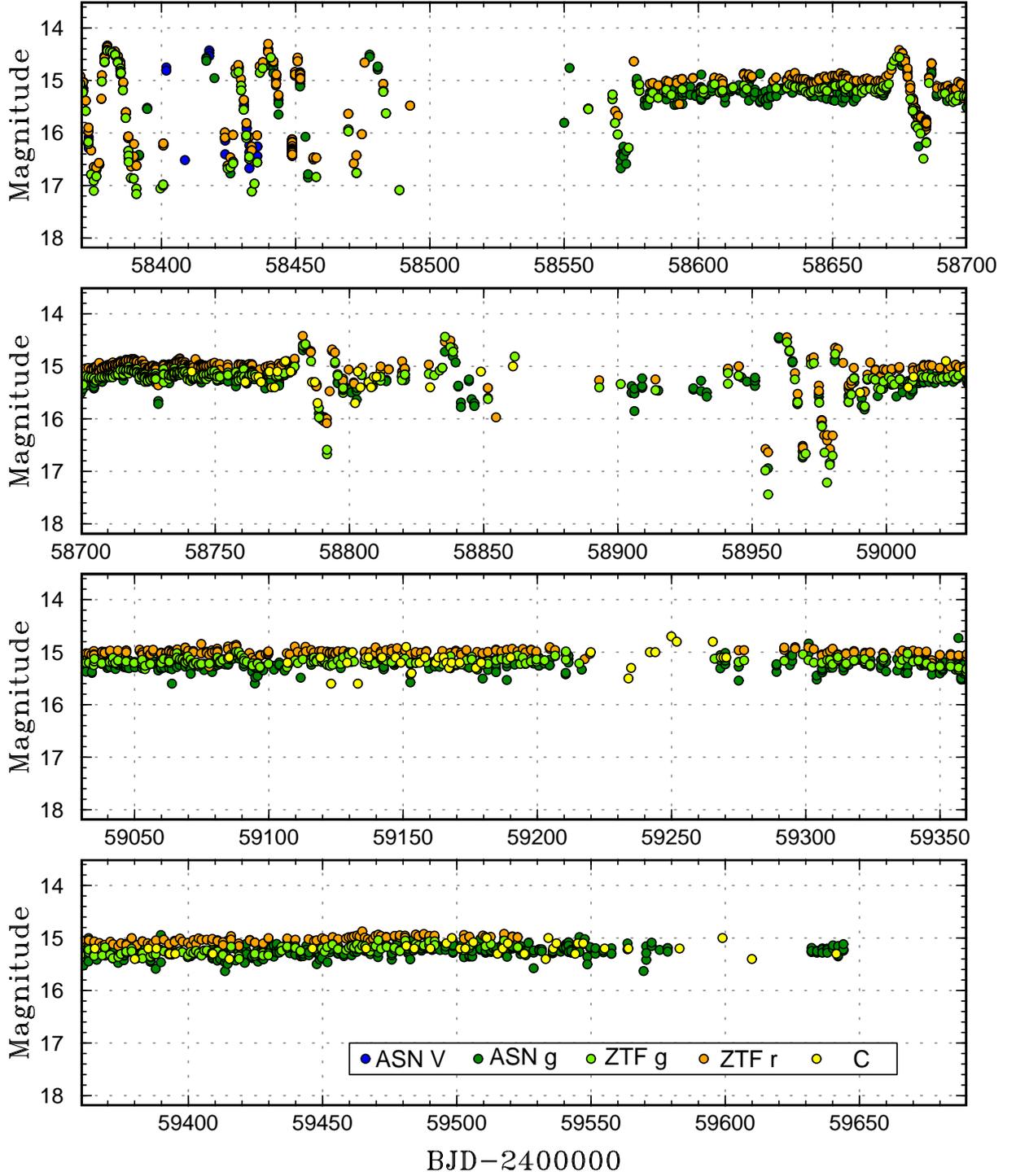}
\caption{
  Long-term light curve of V507 Cyg using
  ASAS-SN, ZTF and our unfiltered CCD observations (C).
  ASN in the legend stands for ASAS-SN.
  This figure is the continuation of figure \ref{fig:v507cyglc}.
  V507 Cyg initially showed dwarf nova-type variations.
  Long standstills then appeared (BJD 2458580),
  which were interrupted
  by brightening followed by a dip and subsequent damping
  oscillations.  After BJD 2458990, this object entered
  a long-lasting standstill, which has continued up to now.
}
\label{fig:v507cyglc2}
\end{center}
\end{figure*}

   When \citet{kat19iwandtype} initially reported on
this object, this object primarily showed dwarf nova-type
variations sometimes interrupted by short standstills,
which had characteristics of IW And stars as follows:
(1) Standstills were terminated by brightening.
(2) There were quasi-periodic cycles
consisting of a (quasi-)standstill with damping oscillations --- 
brightening which terminates the standstill --- a deep dip
and returning to a \mbox{(quasi-)standstill}
(see figure \ref{fig:v507cyglc} around
BJD 2457950--2458040 and BJD 2458275--2458330).
After this IW And-type state, which was followed by an ordinary
dwarf nova state (at the end of figure \ref{fig:v507cyglc}
and the beginning of figure \ref{fig:v507cyglc2}),
long standstills appeared, which were interrupted
by brightening (BJD 2458675, 2458783 and 2458836)
followed by a dip and subsequent damping
oscillations, which are characteristic to an IW And-type
object.  After BJD 2458990 (2020 May), this object entered
a long-lasting standstill, which has continued up to now
(figure \ref{fig:v507cyglc2}).
It would be interesting to note that the standstill
BJD 2458890--2458950 was terminated by fading
as in usual Z Cam stars.

   Such long-lasting standstills had been recorded in
IM Eri \citep{kat20imeri}, BO Cet \citep{kat21bocet}
and ASAS J071404$+$7004.3 \citep{kat22j0714}: all these
objects were initially considered as novalike stars because
of this \citep{che01ECCV,arm13aqmenimeri,
rod07newswsex,ini22j0714}.
A similar IW~And-type object HO Pup
\citep{kim20iwandmodel,lee21hopup}, however, did not
show such a long standstill lasting for a year.

\begin{figure*}
  \begin{center}
    \includegraphics[width=16cm]{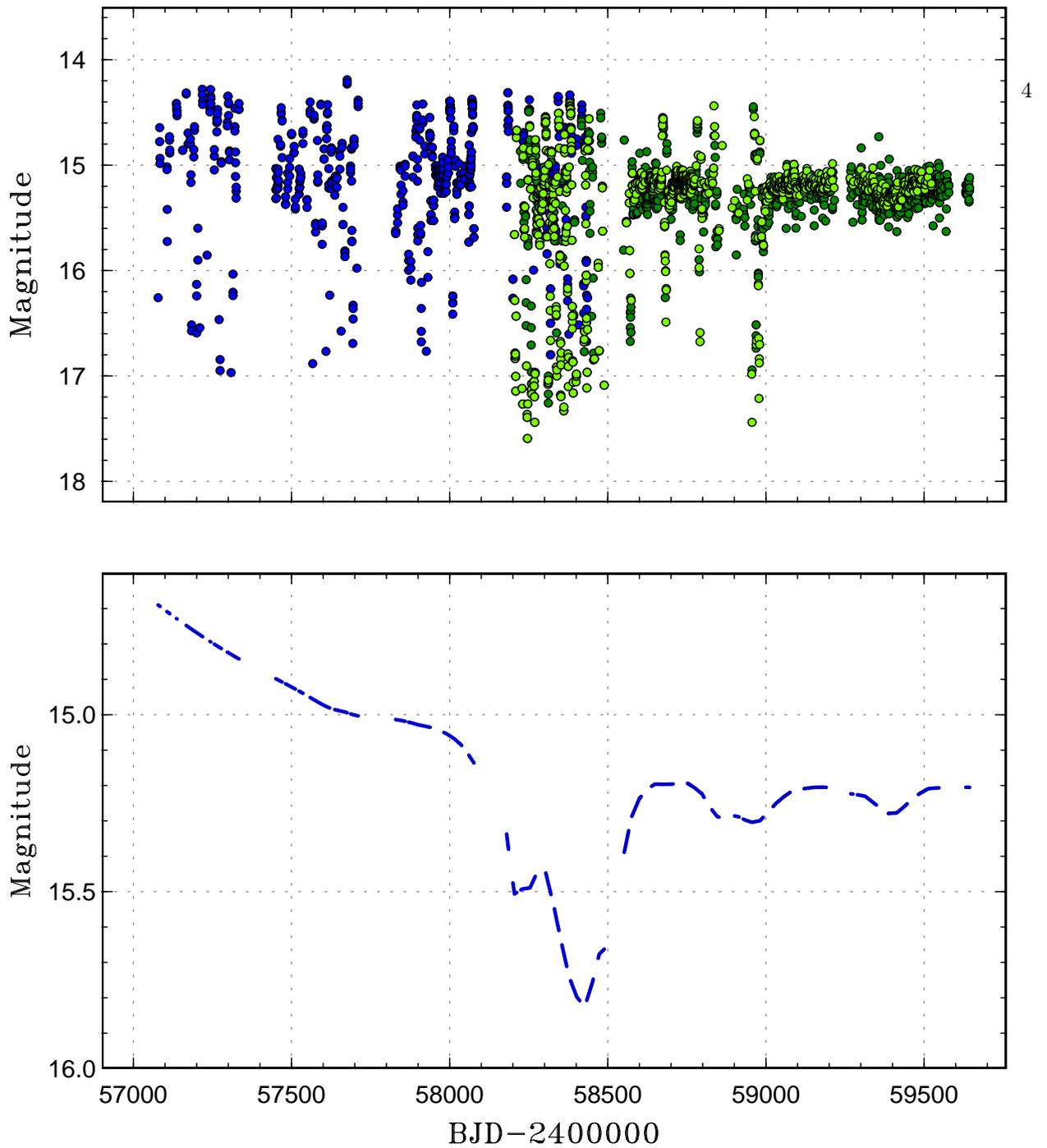}
  \end{center}
  \caption{(Upper): Combined ASAS-SN and ZTF $g$ light curve of
  V507 Cyg.  See figure \ref{fig:v507cyglc2} for the symbols.
  (Lower): Trend determined by LOWESS.  A smoothing parameter
  of $f$=0.20 was used.
  The dwarf nova state (including the IW And-type state)
  in 2018 (BJD 2458200--2458500) corresponded to
  a state fainter than the subsequent state by 0.3~mag
  in average.  Note that the trend is overestimated
  (brighter than real) before BJD 2458000.
  }
  \label{fig:v507cygave}
\end{figure*}

   We analyzed the the trend of mean magnitudes
(averaged in flux) as we performed for two IW And stars
ST~Cha \citep{kat21stchaporb} and
ASAS J071404$+$7004.3 \citep{kat22j0714} using
locally-weighted polynomial regression (LOWESS: \cite{LOWESS}).
The result is shown in figure \ref{fig:v507cygave}.
We should note that the trend before BJD 2458000
was overestimated (brighter than actual) when the ASAS-SN
observations were not deep enough to record
the object at minimum brightness.
The trend after BJD 2458000 can be considered
as real.

   It is apparent that the object was systematically
fainter by 0.3~mag between BJD 2458200 and 2458500,
when the object showed dwarf nova-type variations
including the IW And-type states.
Although it is widely believed that standstills in Z Cam
stars reflect the states with higher mass-transfer
rates \citep{mey83zcam}, the results by observations
have not been very clear
\citep{hon98zcam,kat19nyser,kat21ixvel,kat22j0714}.
The present case provides the first clear evidence
that a dwarf nova-type phase (including the IW And state)
was associated with a fading episode by 0.3~mag 
(corresponding to a 25\% decrease in the mass-transfer rate)
lasting for $\sim$300~d.
The 2019 state with standstills interrupted by brightening
followed by damping oscillations (BJD 2458560--2458860)
was as bright as the following long standstill.
This implies that occasional IW And-type phenomena
also occurred at the same mass-transfer rate as
in the long standstill and that the condition whether
IW And-type phenomena occur or not is subtle.

\section*{Corrections}

\begin{figure*}
  \begin{center}
    \includegraphics[width=16cm]{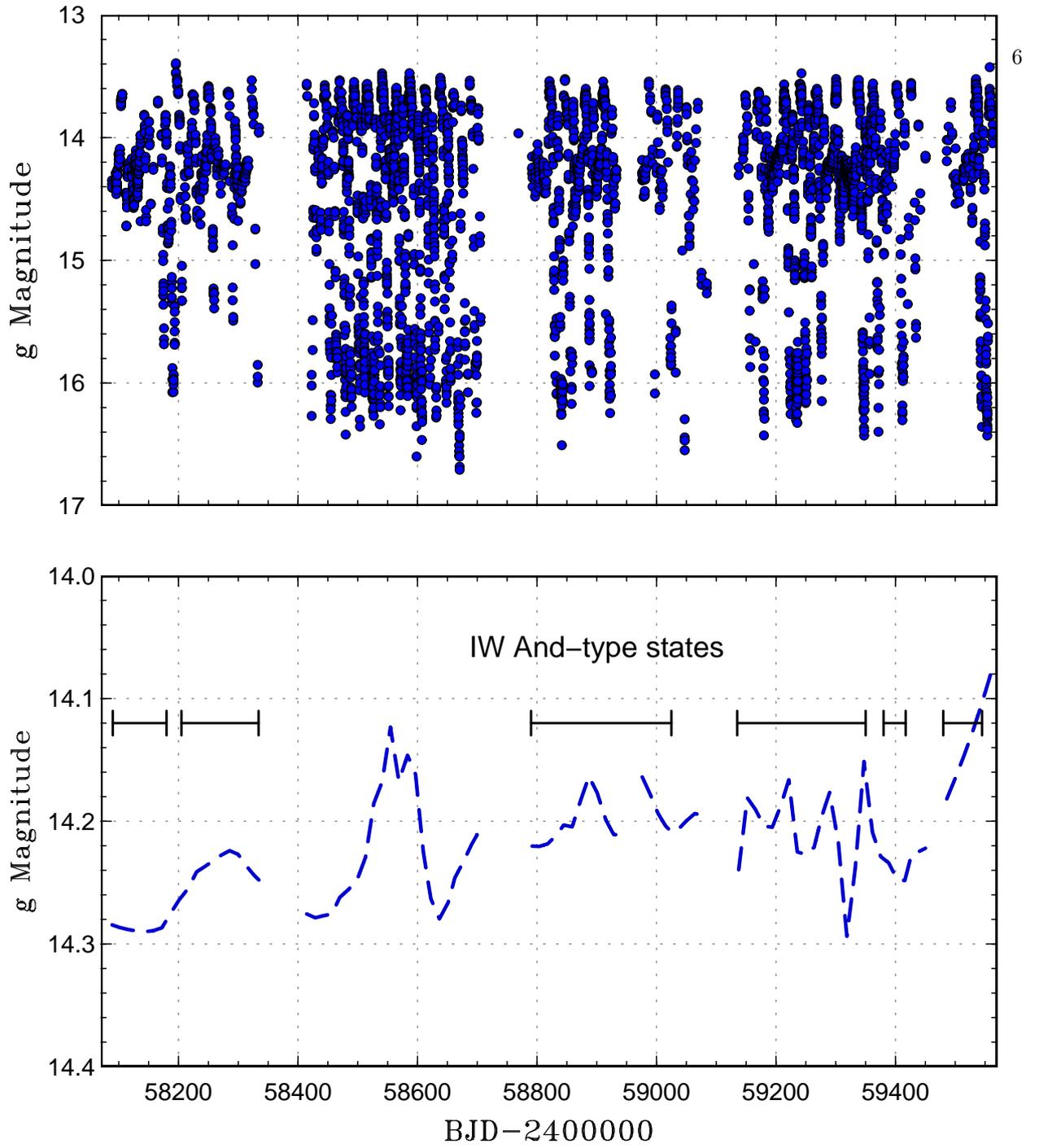}
  \end{center}
  \caption{(Upper): ASAS-SN $g$-band light curve of ST Cha.
  (Lower): Trends determined by LOWESS.
  Horizontal marks represent IW And-type states.
  (Correction of \cite{kat21stchaporb})
  }
  \label{fig:stchaave}
\end{figure*}

\begin{figure*}
  \begin{center}
    \includegraphics[width=16cm]{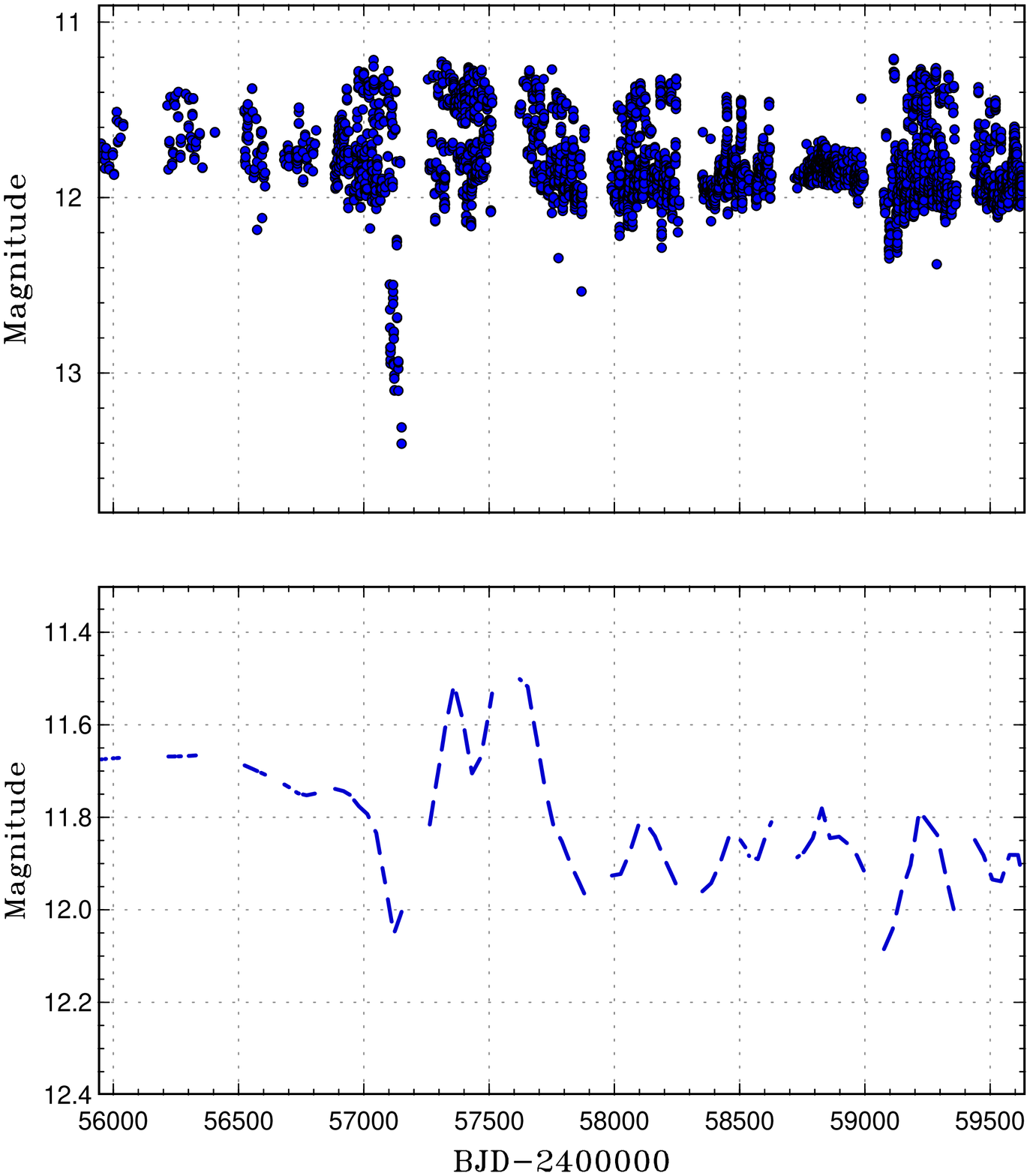}
  \end{center}
  \caption{(Upper): ASAS-SN light curve of ASAS J071404$+$7004.3.
  (Lower): Trend determined by LOWESS.  A smoothing parameter
  of $f$=0.05 was used.
  (Correction of \cite{kat22j0714})
  }
  \label{fig:j0714ave}
\end{figure*}

   While working on this object, TK noticed that there were
scaling errors in ST~Cha \citep{kat21stchaporb} and
ASAS J071404$+$7004.3 \citep{kat22j0714}.
I here present the corrected figures
\ref{fig:stchaave} and \ref{fig:j0714ave}.

\section*{Acknowledgements}

This work was supported by JSPS KAKENHI Grant Number 21K03616.
The author is grateful to the ASAS-SN and ZTF teams
for making their data available to the public.
We are grateful to Naoto Kojiguchi for
helping downolading the ZTF data.

Based on observations obtained with the Samuel Oschin 48-inch
Telescope at the Palomar Observatory as part of
the Zwicky Transient Facility project. ZTF is supported by
the National Science Foundation under Grant No. AST-1440341
and a collaboration including Caltech, IPAC, 
the Weizmann Institute for Science, the Oskar Klein Center
at Stockholm University, the University of Maryland,
the University of Washington, Deutsches Elektronen-Synchrotron
and Humboldt University, Los Alamos National Laboratories, 
the TANGO Consortium of Taiwan, the University of 
Wisconsin at Milwaukee, and Lawrence Berkeley National Laboratories.
Operations are conducted by COO, IPAC, and UW.

The ztfquery code was funded by the European Research Council
(ERC) under the European Union's Horizon 2020 research and 
innovation programme (grant agreement n$^{\circ}$759194
-- USNAC, PI: Rigault).

\section*{List of objects in this paper}
\xxinput{objlist.inc}

\section*{References}

We provide two forms of the references section (for ADS
and as published) so that the references can be easily
incorporated into ADS.

\renewcommand\refname{\textbf{References (for ADS)}}

\newcommand{\noop}[1]{}\newcommand{\hyphalt}{-}

\xxinput{v507cygstaph.bbl}

\renewcommand\refname{\textbf{References (as published)}}

\xxinput{v507cygst.bbl.vsolj}

\end{document}